# 基于中国 A 股交易数据的
# 资本资产定价模型实证研究

任凯

## 摘　要

本文使用中国 A 股市场 2000 年到 2019 年的交易数据，对资本资产定价模型进行实证研究。首先，采用 Fama-MacBetch 回归对标准 CAPM 进行检验，尽管结果成功验证了三个核心假设，但是得到的 $\beta$ 风险对收益率影响并不显著。其次，又分析了 Fama-French 三因子模型，该模型综合使用市场因子、规模因子和价值因子进行资本资产定价，结果显示其能够捕捉到 A 股市场收益率的绝大部分变动，对所构建 25 个投资组合回归得到调整后 $R^2$ 均大于 0.88。最后，本文考虑了中国股市特有的 IPO 监管造成的"壳价值污染"问题，通过剔除市值最低 30%部分的股票来尽量削减其影响，这使得三因子模型中的部分异常结果得到有效修正。尽管本文未提出较为创新的方法，但在数据选取上较为独特，且对数据处理与回归分析过程做了细致呈现，这 1）阐释了资本资产定价模型在中国市场中的适用情况；2）为资本资产定价模型的基础学习提供了开源资料。

**关键词**：中国 A 股市场交易数据　资本资产定价模型　Fama-French 三因子模型

# 一、 资本资产定价模型概述

## 1.1 标准 CAPM

二十世纪六、七十年代，Sharpe（1964），Lintner（1965）和 Black（1972）将 Markowitz（1959）的理论延伸成为资本资产定价模型（Capital and Asset Pricing Model，CAPM），标准 CAPM 用方程表示为：

$$E\left(\widetilde{R}_i\right) = R_f + \left[E\left(\widetilde{R}_m\right) - R_f\right]\beta_i \tag{1}$$

其中 $E\left(\widetilde{R}_i\right)$ 是资产 $i$ 的期望收益率；$R_f$ 指无风险利率；$E\left(\widetilde{R}_m\right)$ 为市场组合的期望收益率，它在理论上是指由所有的风险资产共同构成的投资组合；$\beta_i$ 表示资产 $i$ 的系统风险，它被定义为资产收益与市场组合收益之间的协方差同市场组合收益方差之间的比：

$$\beta_i = \frac{cov\left(\widetilde{R}_i, \widetilde{R}_m\right)}{\sigma^2\left(\widetilde{R}_m\right)}$$

标准 CAPM 是在均值方差假设、投资者一致假设、完全市场假设等严格的假设条件下进行的理论分析模型，其主要结论是：任何资产的期望收益率与其 $\beta$ 值均呈线性正相关，如果市场达到均衡，市场上的所有资产的风险收益定价关系都应在证券市场线上。Black、Jensen 和 Scholes（1972）以及 Fama 和 MacBetch（1973）均证实了股票的平均收益和其 $\beta$ 间存在着显著的正向线性关系。

## 1.2 Fama-Franch 三因子模型

二十世纪后期，一些实证分析结果向标准 CAPM 发起挑战，其中最引人瞩目的就是 Banz（1981）提出的"规模效应"，他发现由股票市值（ME，股票的价格乘以流通股数目）所反映的股票规模能够提高 $\beta$ 值对平均收益率的横截面解释能力。

Fama 和 French（1992a）也和 Ringanum（1981）以及 Lakonishok 和 Shapiro（1986）一样，都发现了 1963 至 1990 年期间，美国股票市场风险 $\beta$ 和平均收益之间的线性正相关关系消失了。他们注意到：股票平均收益率与规模、杠杆、E/P（盈利收益率）以及账面市值比之间的单变量关系很强。在多变量测试中，规模和平均收益率之间的负相关关系是稳健的，账面市值比与平均收益率之间的正向关系也持续存在，进而认为如果资产定价是理性的，那么股票风险应该是多维度的，风险的一个维度是由规模来代表，另一个维度由账面市值比来代表。

实证结果表明规模（ME）和账面市值比（BE/ME）为 1963 至 1990 年期间的股票收益率提供了简单而有效的解释，在此基础上 Fama 和 French（1993）提出了"Fama-French 三因子模型"：

$$R(t) - RF(t) = a + b[RM(t) - RF(t)] + sSMB(t) + hHML(t) + e(t) \quad (2)$$

其中 $t$ 是表示时期 $t$，$R(t)$、$RF(t)$ 和 $RM(t)$ 分别表示 $t$ 时期投资组合的收益率、市场的无风险利率以及市场组合的收益率。$SMB(t)$、$HML(t)$ 为在标准 CAPM（1）的基础上增加的两个因子，其含义分别是 $t$ 时期"小市值组合收益率减大市值组合收益率"以及"高账面市值比组合收益率减低账面市值比组合收益率"。该模型表明：规模和账面市值比这两个因子可以有效解释不同投资组合平均收益的差异。

本文将基于中国 A 股市场数据对标准 CAPM 和 Fama-French 三因子模型进行实证分析，后续章节按照如下步骤展开：第二节将说明本文的数据来源；第三节将介绍 Fama-MacBatch 回归方法以及使用该方法对标准 CAPM 进行检验的结果；第四节则紧接着构建了 Fama-French 三因子模型并展开分析；第五节参考"中国版三因子模型"通过剔除市值最低 30%部分的股票尽量削减"壳价值污染"的影响，进而对三因子模型的部分异常现象做出解释；第六节总结了本文的相关结论。

## 二、 数据来源

本文 A 股的交易数据和指数的交易数据均来自 RESSET 数据库，无风险利率采用一年期定期存款利率，数据频率为月度，时间区间为 2000 年 1 月至 2019 年 12 月。

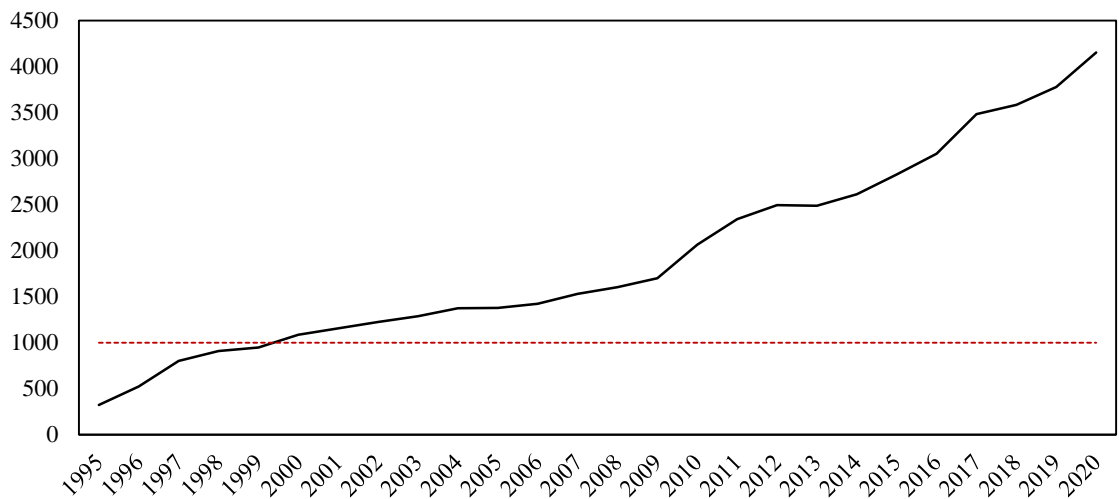

**图 1.** 1995 年至 2020 年中国上市公司数量

中国 A 股交易市场，开始于 1990 年上交所和深交所成立。本文将研究区间截取为 2000 年至 2019 年，主要由于以下三点原因。首先，为了**保证会计数据的统一性：**在中国，有关财务报告各方面的规则和条例直到 1999 年才基本成型。虽然 1993 年规定了公平交易和财务披露原则，但企业在遵守这些原则方面得到的指导较少，各公司自由发挥、强加标准，统一的会计准则在 1999 年才得到广泛实现，因此 2000 年及之后年份的交易样本所依赖的会计数据更具统一性、可比性。其次，为了**确保数据充足：**如图 1.所示，中国上市公司数量 2000 年突破 1000 家，且在此之后产生了井喷式的增长，这样能够充分保证交易数据的数量。最后，考虑到 2020 年以来**新冠疫情对市场产生了较大冲击**，本文暂时选择将 2020 年以后的数据进行剔除。

# 三、 采用 Fama-MacBatch 回归检验标准 CAPM

## 3.1 假设

方程（1）所表示的标准 CAPM 有着 3 个需要被验证的基础条件。**C1**：对任意投资组合而言，预期收益与其风险之间的关系应是线性的；**C2**：由于方程（1）中没有出现证券 $i$ 的其他风险度量因素，那么 $\beta_i$ 应该是该证券风险的完整度量；**C3**：在真实的市场中，较高的风险应与较高的预期收益相匹配，即 $E(\widetilde{R}_m) - E(\widetilde{R}_0) > 0$。

### 3.1.1 收益率的随机模型

为了能够使用观察到的平均收益率数据来验证 **C1~C3**，必须在方程（1）的基础上构造一个逐期收益形式的模型。本文采用与 Fama 和 MacBetch（1973）相同的方法对方程（1）做如下形式的随机概括：

$$\widetilde{R}_{i,t} = \widetilde{\gamma}_{0t} + \widetilde{\gamma}_{1t}\beta_i + \widetilde{\gamma}_{2t}\beta_i^2 + \widetilde{\gamma}_{3t}s_i + \widetilde{\eta}_{i,t} \tag{3}$$

其中，$t$ 指的是时期 $t$；$\widetilde{R}_{i,t}$ 是证券 $i$ 从 $t-1$ 期到 $t$ 期的单期百分比收益率。方程（3）允许 $\widetilde{\gamma}_{0t}$ 和 $\widetilde{\gamma}_{1t}$ 在不同期进行随机变化。注意到，引入 $\beta_i^2$ 是为了检验线性，条件 **C1** 所对应的假设为 $E(\widetilde{\gamma}_{2t}) = 0$。类似的，方程（3）中引入的 $s_i$ 是对证券 $i$ 某种未能被 $\beta_i$ 度量的风险的衡量，因此条件 **C2** 所对应的假设应为 $E(\widetilde{\gamma}_{3t}) = 0$。条件 **C3** 对应于方程（3）中风险溢价 $\widetilde{\gamma}_{1t}$ 的期望为正，即 $E(\widetilde{\gamma}_{1t}) > 0$。当然在对比方程（1）的和方程（3）后，还可以确定一个显而易见需要被检验的假设 $E(\widetilde{\gamma}_{0t}) = R_{ft}$。

### 3.1.2 模型的核心假设

总的来说，本文针对标准 CAPM 其给出如下四个可检验的核心假设：

**H1（线性）**：$E(\tilde{\gamma}_{2t}) = 0$；

**H2（β 是风险的完整度量）**：$E(\tilde{\gamma}_{3t}) = 0$

**H3（正预期的收益-风险交易）**：$E(\tilde{\gamma}_{1t}) > 0$

**H4（Sharpe-Lintner 假设）**：$E(\tilde{\gamma}_{0t}) = R_{ft}$

## 3.2 检验方法

### 3.2.1 方法概述

检验标准 CAPM 不得不面对变量误差（errors-in-variables, EIV）问题，即在实际操作中只能使用风险度量的估计值$\hat{\beta}_i$而非方程（3）所要求的真实值$\beta_i$。本文采用$\hat{\beta}_i$的定义式进行估计：

$$\hat{\beta}_i = \frac{\widehat{cov}(\widetilde{R}_i, \widetilde{R}_m)}{\hat{\sigma}^2(\widetilde{R}_m)}$$

其中$\widehat{cov}(\widetilde{R}_i, \widetilde{R}_m)$和$\hat{\sigma}^2(\widetilde{R}_m)$分别是使用月度收益数据对$cov(\widetilde{R}_i, \widetilde{R}_m)$和$\sigma^2(\widetilde{R}_m)$的估计。本文选用上证综合指数作为对市场组合的近似替代，其样本股是在上海证券交易所上市的所有股票并通过派许加权综合价格指数公式完成计算。选用上证综合指数一方面是考虑到其是中国的大股指能够较好地反映市场变动，另一方面是因为数据区间匹配[1]。

Blume（1970）证明对于任何投资组合 $p$，如果其是多支证券基于权重 $x_{ip}$，$i = 1, 2, ..., N$ 来定义的，那么投资组合 $p$ 的风险就可以表示为：

---

[1] 上证综合指数于 1990 发布，完整地覆盖了 2000 年到 2019 年整个数据区间

$$\hat{\beta}_p = \frac{\widehat{cov}\left(\widetilde{R}_p, \widetilde{R}_m\right)}{\hat{\sigma}^2\left(\widetilde{R}_m\right)} = \sum_{i=1}^{N} x_{ip} \frac{\widehat{cov}\left(\widetilde{R}_i, \widetilde{R}_m\right)}{\hat{\sigma}^2\left(\widetilde{R}_m\right)} = \sum_{i=1}^{N} x_{ip} \hat{\beta}_i$$

经过严谨数学推导后不难发现：投资组合的$\hat{\beta}_p$往往比单支证券的$\hat{\beta}_i$更贴近真实的$\beta$。同时，为了减少因使用投资组合造成的信息损失，一般是在对单支证券$\hat{\beta}_i$排序的基础上构建投资组合，然后通过平均或市值加权的方法获得投资组合的$\hat{\beta}_p$值。但需要注意的是，如果只是简单地执行上述操作很可能会导致严重的回归现象。在基于单个证券回归出的$\hat{\beta}_i$截面上，较大的$\hat{\beta}_i$往往高于相应的真实$\beta_i$，较小的$\hat{\beta}_i$往往低于真实$\beta_i$，而在对$\hat{\beta}_i$排序的基础上构建投资组合会造成抽样误差的堆积。为了减小回归现象造成的不利影响，本文采用 Fama 和 MacBetch（1973）提出的"Fama-MacBatch 回归"完成投资组合的构建以及后续回归的计算，具体方式如下：

**步骤 1.** 将基于一个时间段内交易数据计算出的单支股票$\hat{\beta}_i$进行排序，按照$\hat{\beta}_i$由小到大构建出 20 个投资组合，该时间段被称为"组合构建期"；

**步骤 2.** 用随后时间段中的数据，分别估计出 20 个投资组合中每支股票的初始$\hat{\beta}_i$，然后在组合内部平均，将这段时间称为"初始估计期"；

**步骤 3.** 将初始估计期按年向前滚动，分别估计 20 个投资组合后续每一年的$\hat{\beta}_p$以及每月的平均收益率，进而完成对模型的回归检验，将这段时间称为"模型检验期"。

这样，使用非本期时段中的"新"数据能够保证一个投资组合中单支股票$\hat{\beta}_i$的估计误差在很大程度上是随机的，进而使估计投资组合$\hat{\beta}_p$时所产生回归现象的不利影响被尽量减弱。

### 3.2.2 方法细节

图 2.以 2000 到 2012 年数据区间为例，展示了本文进行参数估计的细节。首先使用前 4 年（2000~2003）"组合构建期"的月度收益数据估计出每一股票的$\hat{\beta}_i$，按照由小到大的顺序均分出 20 个投资组合[2]。

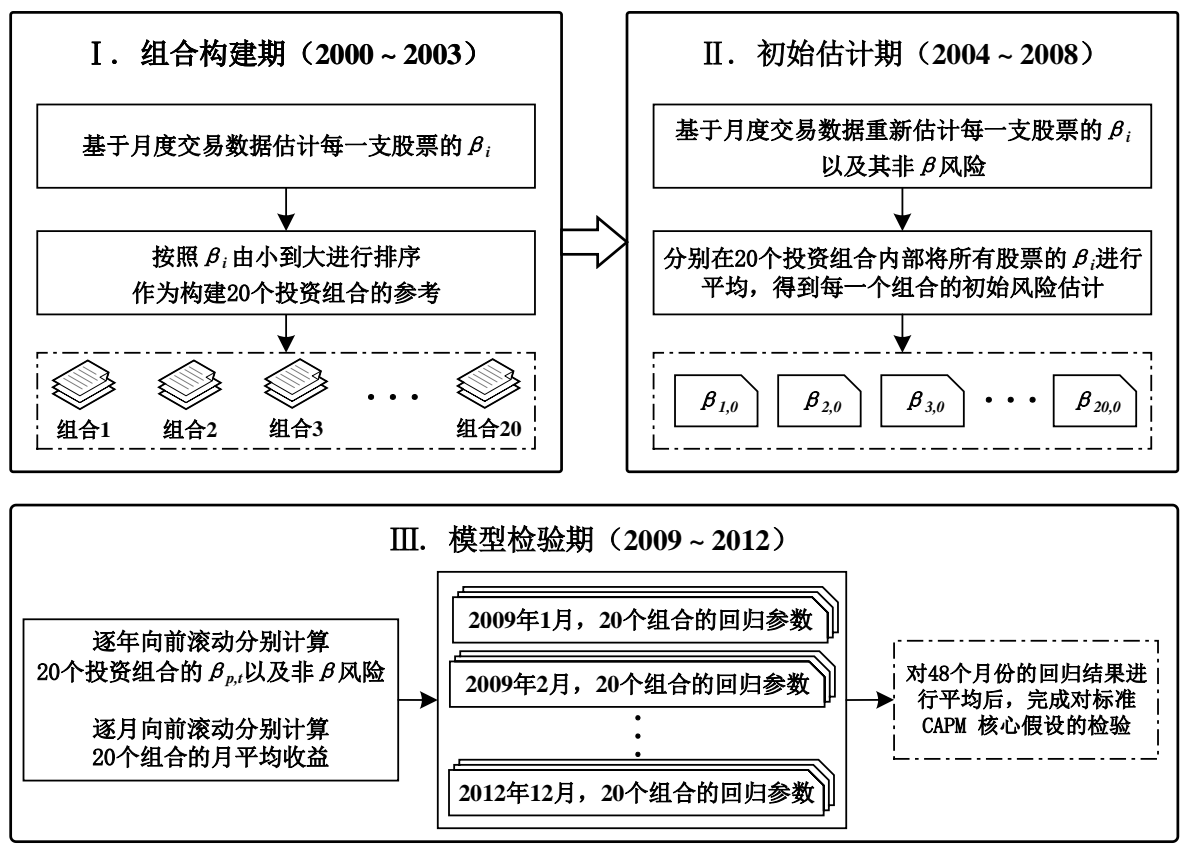

**图 2.** 采用 Fama-MacBatch 回归检验标准 CAPM 方法框架图
（以 2000 到 2012 年数据区间为例）

随后 5 年（2002~2008 年）"初始估计期"的数据被用来重新估计个股的$\hat{\beta}_i$，分别在已构建的 20 个投资组合内对个股的$\hat{\beta}_i$取均值得到组合的初始风险估计$\hat{\beta}_{p,0}$，在接下来的 4 年（2009~2011 年）中每年都会重新**滚动计算**$\hat{\beta}_{p,t}$。注意到，本文和 Fama 和 MacBetch（1973）所用的方法并不完全一致，其对$\hat{\beta}_{p,t}$进行的是**逐月滚动**调整，但本文考虑到风险波动的周期性，$\hat{\beta}_{p,t}$变动不会过于频繁，故采用**逐年滚动**的方法进行调整。

---

[2] 1 号投资组合到 20 号投资组合中股票的$\hat{\beta}_i$递减。

本文使用 $s(\hat{\epsilon}_i)$ 作为对股票 $i$ 非 $\beta$ 风险的衡量，其含义为在估计市场模型（4）时所产生的最小二乘残差 $\hat{\epsilon}_{it}$ 的标准差：

$$\widetilde{R}_{it} = \alpha_i + \beta_i \widetilde{R}_{mt} + \tilde{\epsilon}_{it} \tag{4}$$

这样做是因为 $\beta_i = cov(\widetilde{R}_i, \widetilde{R}_m)/\sigma^2(\widetilde{R}_m)$，在（4）中有 $cov(\tilde{\epsilon}_i, \widetilde{R}_m) = 0$，进而有：

$$\sigma^2(\widetilde{R}_i) = \beta_i^2 \sigma^2(\widetilde{R}_m) + \sigma^2(\tilde{\epsilon}_i) + 2\beta_i cov(\widetilde{R}_m, \tilde{\epsilon}_i)$$

显然，$s(\hat{\epsilon}_i)$ 是对股票 $i$ 收益分布中与 $\beta_i$ 不直接相关的分散性的估计，也即股票 $i$ 的非 $\beta$ 风险。

在 2009～2012 年的 4 年间，20 个投资组合的**逐月平均收益率** $R_{p,m}$ 也被计算出来。对于这一时期所有月份的数据，进行如下截面回归：

$$R_{p,m} = \hat{\gamma}_{0m} + \hat{\gamma}_{1m}\hat{\beta}_{p,y-1} + \hat{\gamma}_{2m}\hat{\beta}_{p,y-1}^2 + \hat{\gamma}_{3m}\bar{s}_{p,y-1} + \hat{\eta}_{p,m} \tag{5}$$

$\hat{\beta}_{p,t-1}$ 是组合 $p$ 中所有股票 $\hat{\beta}_i$ 的平均值；$\hat{\beta}_{p,t-1}^2$ 是组合中 $\hat{\beta}_i$ 的平方的平均值；$\bar{s}_{p,t-1}(\hat{\epsilon}_i)$ 是组合中每一支股票 $s(\hat{\epsilon}_i)$ 的平均值；$m$ 表示月度，$y$ 表示年度。

方程（5）是（2）对一个投资组合中所有股票平均化后的形式，把逐年的 $\hat{\beta}_{p,t-1}$，$\hat{\beta}_{p,t-1}^2$，以及 $\bar{s}_{p,t-1}(\hat{\epsilon}_i)$ 作为解释变量，逐月的平均收益率作为被解释变量，对随机系数 $\hat{\gamma}_{0m}$，$\hat{\gamma}_{1m}$，$\hat{\gamma}_{2m}$ 以及 $\hat{\gamma}_{3m}$ 进行最小二乘估计。使用 2009~2012 这 4 年间的数据对方程（5）进行回归得到的逐月系数估计值是该时期标准 CAPM 模型假设检验的依据，要想获得其他时期的检验结果，只需重复执行上述步骤。

表 1 显示本文基于数据选取的 3 个不同的投资组合构建期（除第一个都是 7 年）、5 年的初始估计期和检验期（除了最后一个都是 4 年）。在对不满足要求的数据进行剔除时，本文直接选取了 2000 年到 2019 年 20 年里均有交易记录的共 724 支股票数据，尽管其中有部分股票个别月份停止交易，但采用上述提到的逐年滚动的方法可以减弱其影响。

表 1. 三组"组合构建期"、"初始估计期"、"模型检验期"

|  | 时期 | | |
| --- | --- | --- | --- |
|  | 1 | 2 | 3 |
| 组合构建期 | 2000~2003 | 2001~2007 | 2005~2011 |
| 初始估计期 | 2004~2008 | 2008~2012 | 2012~2016 |
| 模型检验期 | 2009~2012 | 2013~2016 | 2017~2019 |

### 3.2.3 方法实施示例

为了便于读者理解，此处以表 1.中的时期 1（2000 至 2012 年）为例，展示本方法具体实施步骤。

**步骤 1.** 基于"组合构建期"2000 年至 2003 年中的数据，采用定义式估计出每一支股票的 $\hat{\beta}_i$：

$$\left(\hat{\beta}_1, \hat{\beta}_2, \cdots, \hat{\beta}_{724}\right)'$$

对所有股票的 $\hat{\beta}_i$ 进行排序，由小到大等分地构建起 20 个投资组合：$\hat{\beta}_i$ 最小 5%对应的股票为投资组合 1（38 支股票），其次为投资组合 2（36 支股票）以此类推，最大的 5%对应的证券为投资组合 20（38 支股票）。

**步骤 2.** 根据"初始估计期"中 60 个月的数据，用市场模型（4）估计出该区间中每一支股票的 $\hat{\beta}_i$ 以及 $s(\hat{\epsilon}_i)$，按照上一步得到的投资组合划分，在 20 个投资组合内部对每一支股票进行平均进而获得每个投资组合在初期的 $\hat{\beta}_{p,0}$ 以及 $\bar{s}_{p,0}$：

$$\hat{\beta}_{p,0} = \left(\hat{\beta}_{p1,0}, \hat{\beta}_{p2,0}, \cdots, \hat{\beta}_{p20,0}\right)'$$
$$\bar{s}_{p,0} = \left(\bar{s}_{p1,0}, \bar{s}_{p2,0}, \cdots, \bar{s}_{p20,0}\right)'$$

**步骤 3.** 计算"模型检验期"第一年里 12 个月 20 个投资组合的平均收益率：

$$R_{p,1} = \begin{bmatrix} R_{p1,1} & \cdots & R_{p20,1} \\ \vdots & \ddots & \vdots \\ R_{p1,12} & \cdots & R_{p20,12} \end{bmatrix}$$

向前滚动 1 年，基于 2004 年至 2009 年 72 个月的数据，重复上述操作，得到当期 $\hat{\beta}_{p,1}$ 以及 $\bar{s}_{p,1}$：

$$\hat{\beta}_{p,1} = \left(\hat{\beta}_{p1,1}, \ \hat{\beta}_{p2,1}, \ \cdots, \ \hat{\beta}_{p20,1}\right)'$$

$$\bar{s}_{p,1} = \left(\bar{s}_{p1,1}, \ \bar{s}_{p2,1}, \ \cdots, \ \bar{s}_{p20,1}\right)'$$

紧接着计算"模型检验期"第二年 2010 年 12 个月 20 个投资组合的平均收益率

$$R_{p,2} = \begin{bmatrix} R_{p1,13} & \cdots & R_{p20,13} \\ \vdots & \ddots & \vdots \\ R_{p1,24} & \cdots & R_{p20,24} \end{bmatrix}$$

继续向前滚动 1 年，基于 2004 年至 2010 年 84 个月的数据，重复上述操作，得到当期 $\hat{\beta}_{p,2}$、$\bar{s}_{p,2}$ 以及 $R_{p,3}$；进而基于 2004 年至 2011 年 96 个月的数据，重复上述操作，得到当期 $\hat{\beta}_{p,3}$、$\bar{s}_{p,3}$ 以及 $R_{p,4}$。汇总 2009 年至 2012 年 4 年的所有数据，得到如表 2. 所示的数据结构，为后续回归估计做准备。

**表 2.** 数据预处理结果（20 个投资组合在 48 个月份上的截面数据）

| 组合 | 时间 | $R_p$ | $\hat{\beta}_p$ | $\hat{\beta}_p^2$ | $\bar{s}_p$ |
|---|---|---|---|---|---|
| | 2009 年 1 月 | $R_{p1,1}$ | $\hat{\beta}_{p1,0}$ | $\hat{\beta}_{p1,0}^2$ | $\bar{s}_{p1,0}$ |
| 组合 1 | $\vdots$ | $\vdots$ | $\vdots$ | $\vdots$ | $\vdots$ |
| | 2012 年 12 月 | $R_{p1,48}$ | $\hat{\beta}_{p1,3}$ | $\hat{\beta}_{p1,3}^2$ | $\bar{s}_{p1,3}$ |
| $\vdots$ | $\vdots$ | $\vdots$ | $\vdots$ | $\vdots$ | $\vdots$ |
| | 2009 年 1 月 | $R_{p20,1}$ | $\hat{\beta}_{p20,0}$ | $\hat{\beta}_{p20,0}^2$ | $\bar{s}_{p20,1}$ |
| 组合 20 | $\vdots$ | $\vdots$ | $\vdots$ | $\vdots$ | $\vdots$ |
| | 2012 年 12 月 | $R_{p20,48}$ | $\hat{\beta}_{p20,3}$ | $\hat{\beta}_{p20,3}^2$ | $\bar{s}_{p20,3}$ |

## 3.2 估计结果与假设检验

### 3.2.1 估计结果呈现

本文对全样本区间（2009～2019）和三个子样本区间（2009～2012、2013～2016、2017～2019）分别做了分析，表 3.呈现了四个时间区间下相关参数的估计结果。本文分四组对方程（5）的四种形式进行回归，D 组是基于（5）本身。A 到 C 组里（5）中的一个或多个变量被移除。表格中显示的 $\overline{\hat{\gamma}}_j$ 是逐月回归系数 $\hat{\gamma}_{jt}$ 的平均值；$s(\hat{\gamma}_j)$ 是每月相应估计值的标准差，本表还展示了用于检验 $\overline{\hat{\gamma}}_j = 0$ 这一假设的 $t$ 统计量：

$$t\left(\overline{\hat{\gamma}}_j\right) = \frac{\overline{\hat{\gamma}}_j}{s(\hat{\gamma}_j)/\sqrt{n}}$$

其中 $n$ 是时期内的月份数，也是用于计算 $\overline{\hat{\gamma}}_j$ 和 $s(\hat{\gamma}_j)$ 的 $\hat{\gamma}_{jt}$ 数量。

在解释这些 $t$ 统计量时，本文参考了 Fama（1965a）和 Blume（1970）的结论。这些结论表明，相对于正态分布而言，普通股票的收益分布是"厚尾"的，而且可能更符合非正态对称稳定分布。从 Fama 和 Babiak（1968）的观点来看，这一结论意味着，当人们在基础变量为正态的假设下解释大的 $t$ 统计量时，得到的显著性水平可能是高估的。但需要注意的是，除了假设 *H3*（正的预期收益-风险权衡）之外，向上偏的概率水平会导致产生拒绝标准 CAPM 假设的偏差，**因此如果在正态的假设下解释 $t$ 统计量时不能拒绝这些假设，那么考虑收益分布的厚尾时，这些假设就会更加可信。**

### 3.2.2 核心假设检验

先来看 *H1*：表 3.中 B 组和 D 组的结果**不能显著地拒绝**收益率和 $\beta$ 之间的关系是线性的假设。不难发现，B 组和 D 组之中无论是全样本时间段还是子时间间段内的 $t\left(\overline{\hat{\gamma}}_2\right)$ 均十分接近于 0，尽管子时间段 2017~2019 内的 $t\left(\overline{\hat{\gamma}}_2\right)$ 较大，但也均小于 1.8，仍然无法显著地拒绝原假设。

表 3. 标准 CAMP 参数估计结果汇总

$$R_{p,m} = \hat{\gamma}_{0m} + \hat{\gamma}_{1m}\hat{\beta}_{p,y-1} + \hat{\gamma}_{2m}\hat{\beta}^2_{p,y-1} + \hat{\gamma}_{3m}\bar{s}_{p,y-1} + \hat{\eta}_{p,m}$$

| 时间区间 | $\overline{\hat{\gamma}_0 - R_f}$ | $t(\overline{\hat{\gamma}_0 - R_f})$ | $\overline{\hat{\gamma}}_1$ | $t(\overline{\hat{\gamma}}_1)$ | $\overline{\hat{\gamma}}_2$ | $t(\overline{\hat{\gamma}}_2)$ | $\overline{\hat{\gamma}}_3$ | $t(\overline{\hat{\gamma}}_3)$ | $R^2$ |
|---|---|---|---|---|---|---|---|---|---|
| A 组 | | | | | | | | | |
| 2009~2019 | .0107 | 1.06 | .0013 | .18 | … | … | … | … | .09 |
| 2009~2012 | .0066 | .32 | .0103 | .70 | … | … | … | … | .04 |
| 2013~2016 | .0306 | 2.02 | -.0070 | -.74 | … | … | … | … | .10 |
| 2017~2019 | -.0071** | -.67 | .0004 | .02 | … | … | … | … | .15 |
| B 组 | | | | | | | | | |
| 2009~2019 | -.0393 | -.21 | .0993 | .27 | -.0477 | -.26 | … | … | .17 |
| 2009~2012 | -.1608 | -.31 | .3268 | .32 | -.1488 | -.30 | … | … | .12 |
| 2013~2016 | .0113 | .12 | .0402 | .22 | -.0270 | -.32 | … | … | .17 |
| 2017~2019 | .0553 | .81 | -.1250 | -.93 | .0595 | .96 | … | … | .22 |
| C 组 | | | | | | | | | |
| 2009~2019 | .0207 | .17 | .0008 | .11 | … | … | -.0738 | -.08 | .17 |
| 2009~2012 | -.2253 | -.13 | .0100 | .74 | … | … | .2254 | .17 | .14 |
| 2013~2016 | .0767 | .31 | -.0091 | -1.02 | … | … | -.3437 | -.18 | .16 |
| 2017~2019 | .0039 | .02 | .0005 | .03 | … | … | .1128 | -.08 | .21 |
| D 组 | | | | | | | | | |
| 2009~2019 | -.0372 | -.14 | .1085 | .27 | -.0528 | -.27 | -.0236 | -.02 | .23 |
| 2009~2012 | -.2823 | -.50 | .3874 | .37 | -.1807 | -.35 | .7314 | .54 | .22 |
| 2013~2016 | .0233 | .04 | .0375 | .13 | -.0248 | -.18 | -.0912 | -.03 | .22 |
| 2017~2019 | .1974 | .82 | -.1686 | -1.14 | .0804 | 1.16 | -.9403 | -.60 | .28 |

$^*p < 0.1, ^{**}p < 0.05, ^{***}p < 0.01$

再考虑标准 CAPM 条件 C2 所对应的假设 **H2**：除了 $\beta$ 以外，没有任何其他风险会系统性地影响预期收益。表 3.中 C 组和 D 组的结果也都**无法显著拒绝**这一假设，可以发现 $t(\overline{\hat{\gamma}}_3)$ 都十分接近于 0，且 $t(\overline{\hat{\gamma}}_3)$ 的正负情况均有出现。

如果只看前面两个假设的话，本文的数据似乎能够很好地印证标准 CAPM 模型的有效性。然而，如果关键条件 C3 对应的假设 *H3* 无法显著被接受，以上得到结论的价值都会大打折扣。也就是说，除非风险和收益之间平均存在正的相关性，否则本文对标准 CAPM 的检验就无法达到令人满意的地步。

遗憾的是，表 3.中的结果和 Fama 和 MacBetch（1973）得到的结果大相径庭，所有模型任意时间区间内的 $t\left(\overline{\hat{\gamma}_1}\right)$ 均比较小，**无法显著拒绝** $\gamma_1 = 0$。然而值得欣慰的是，本文对 *H4* 检验的结果似乎并不像 Fama 和 MacBetch（1973）那样是"ambiguous"的，四组模型中仅有 A 组中两个区间的数据支持本文拒绝 *H4*，其余部分的 $t\left(\overline{\hat{\gamma}_0 - R_f}\right)$ 均十分接近于 0。

## 3.3 对标准 CAPM 检验的结论

综上，本文的结果在一定程度上证实了标准 CAPM 的有效性。一方面，可以说明尽管在不同时期存在"随机非线性"，但是不能拒绝这样的假设：即在做出投资组合决策时，投资者应该认定证券的投资组合风险和其预期收益之间的关系是线性的。同样结果也不支持拒绝"除了投资组合风险外，没有任何风险措施会系统地影响平均收益"这一假设。另外还说明了无风险利率在标准 CAPM 中扮演的重要角色。另一方面不可否认的是，本文无法得到"平均而言，收益和风险之间应该有一个正向的关系"这一重要的核心结论，这很有可能是因为单一的 $\beta$ 因子已经不能够准确反映收益率的变动情况。这也促使本文进一步展开了对 Fama-French 三因子模型的实证研究，想要尝试探索一种更为恰当的资本资产定价模型来解释中国股票市场的收益率变动。

# 四、 Fama-French 三因子模型

## 4.1 数据补充说明

本文将引入新的因子,在此对数据做补充说明。首先,参考 Fama 和 French(1993)的做法排除了金融公司,因为这些公司的部分指标可能与非金融公司的含义不同;其次,剔除了状态为 ST(特别处理)、*ST(退市风险警示)、PT(特别转让)的股票以及账面价值为负的股票,避免这些异常股票对数据产生干扰。

在后续分析中,本文将公司资产负债表中披露 $t-1$ 年末的账面所有者权益作为 $t$ 年**账面价值**,把公司披露的 $t-1$ 年末的市场总值作为 $t$ 年的**市场价值**。同时,不再直接采用上证综合指数作为市场回报率,而是采用流通市值加权平均计算得出。

在我国以往的信息披露规范制度下,上市公司本年度财务报表一般在次年 3、4 月份公布,4 月底为年报披露截止日期,这就造成了财务数据更新存在一定的滞后性,$t$ 年末的财务数据可能直到 $t+1$ 年 4 月份披露财务报表时才能得到更新,进而导致数据库中公司公布的财务数据与实际交易数据时间错位。因此本文**选取 $t$ 年 5 月至 $t+1$ 年 4 月作为组合构建周期**,尽可能减少时间错位带来的影响。例如某只股票母公司在 2002 年 4 月底才披露 2001 年的年度财务报表,那么 2002 年 1 月至 4 月只能依靠 2001 年披露的 2000 年末的账面价值和市场价值数据进行分组,而 2002 年 5 月至 2003 年 4 月可以采用 2002 年 4 月底披露的数据进行分组,以此类推。

## 4.2 时间序列回归的输入

不同于上述参数估计时采用的截面回归,分析 Fama-French 三因子模型需要进行**时间序列回归**。如方程(2)所示,解释变量包括**市场因子**(市场组合的超额收益率)、**规模因子**(市值)和**价值因子**(账面市值比),被解释变量为根据规模(市值)和账

面市值比（$BE/ME$）构建的 25 个股票组合的*平均超额收益率*。本小节将详细介绍这些输入数据的获取方法，并对最终数据结构做展示，之后进行了一些描述性统计。

### 4.2.1 解释变量：股票三因子

*添加因子的动机*——规模和账面市值比是解释股票收益的特别变量，它们能代表除$\beta$因子外的其他基本风险因素。Fama 和 French（1992b）阐述了公司规模和账面市值比与收益率之间的关系，他们发现高账面市值比的公司往往有较高的资产收益，而在控制了账面市值比后，小公司往往比大公司的收益率高。这表明规模与一个市场风险因子有关，该因子可能解释了规模和平均收益率之间的*负相关关系*；而相对盈利能力也是收益率中市场风险的来源，这可能解释了账面市值比和平均收益率的*正相关关系*。三因子模型的核心目的正是通过新增规模因子和价值因子来更好地刻画风险，进而更为准确地解释股票收益率变动情况。

*计算因子的组合构建*——为了研究月度的经济基本面，Fama 和 French（1992b）基于在规模和账面市值比上排序的所有股票构建了六个投资组合。本文采用同样的方法构建投资组合，旨在衡量收益中与规模因子、价值因子相关的的基本风险因素。

表 4. 基于规模和账面市值比联合划分的六种公司类型

|  | 低$BE/ME$（$L$） | 中$BE/ME$（$M$） | 高$BE/ME$（$H$） |
|---|---|---|---|
| 小规模（$S$） | $S/L$ | $S/M$ | $S/H$ |
| 大规模（$B$） | $B/L$ | $B/M$ | $B/H$ |

Fama 和 French（1992a）证明了账面市值比在解释股票收益率上比规模具有更大的作用，因此根据规模将公司分为两类，而根据账面市值比将公司分为三类。表 4.展示了本文划分公司类型的具体方法：在**规模维度**上以$t$年 4 月 A 股所有股票的市值中位数将$t$年 5 月至$t+1$年 4 月"组合构建周期"中的公司分为小规模和大规模（S 和 B）两组；在**价值维度**上按$BE/ME$尾部 30%、中部 40%以及头部 30%三个分位点将公司分为低价值、中价值和高价值三组（L、M 和 H）。联合两大维度就可以构建出六

个投资组合（$S/L$，$S/M$，$S/H$，$B/L$，$B/M$，$B/H$）。$S/L$ 组合包含小规模组中的股票，它们同时也在低价值组中；$B/H$ 组合包含大规模组中的股票，它们也在高价值组中。投资组合在每年 4 月底重新分组[3]。

*规模因子*——用投资组合计算 $SMB(small-big)$ 来测度与规模相关的风险因子，具体计算方法为每个月 3 个小市值投资组合（$S/L$，$S/M$，$S/H$）以及 3 个大市值投资组合（$B/L$，$B/M$，$B/H$）在平均收益率之间的差额：

$$SMB(t) = \frac{R_{S/L}(t) + R_{S/M}(t) + R_{S/H}(t)}{3} - \frac{R_{B/L}(t) + R_{B/M}(t) + R_{B/H}(t)}{3} \qquad (6)$$

将账面市值比作为控制变量，这种差额应该在很大程度上不受账面市值比的影响，而是集中在大规模与小规模公司的收益率差异上。

*价值因子*——类似的，采用 $HML(high-low)$ 衡量与账面市值比相关的风险因子。$HML$ 是每个月里两个高价值组合（$S/H$ 和 $B/H$）平均收益率和两个低价值组合（$S/L$ 和 $B/L$）平均收益率之间的差额：

$$HML(t) = \frac{R_{S/H}(t) + R_{B/H}(t)}{2} - \frac{R_{S/L}(t) + R_{B/L}(t)}{2} \qquad (7)$$

这两个平均收益率之间的差额应该在很大程度上与规模因子无关，而是集中在高价值和低价值公司之间的不同收益率上。

注意，和检验标准 CAPM 不同，此处投资组合的平均收益率不再由组合中个股收益率直接算术平均得到，而是由个股流通市值加权平均计算。这样做一方面是为了最小化方差，另一方面也可以更为准确地模拟真实的投资组合。

*市场因子*——股票收益率的市场因子反映的是超额市场收益率，即 $RM-RF$。其中，$RM$ 不再直接用上证综合指数来代表，而是由个股流通市值加权平均法计算得到，$RF$ 为使用一年期定期存款利率表示的无风险利率。

---

[3] 正如上面所提到的，这是为了确保市值以及账面市值比这两大指标是已知的。

### 4.2.2 被解释变量：收益率

依据规模和账面市值比的排序构建起 25 个投资组合，并将各个组合的平均超额收益率作为时间序列回归的被解释变量。之所以使用按规模和账面市值比排序构建投资组合，是因为要确定$SMB$和$HML$这两个新增因子是否涵盖了与规模、账面市值比相关的股票收益率中的风险。

此处构建 25 个$SIZE-BE/ME$投资组合的方式类似于上文解释的 6 个$SIZE-BE/ME$的投资组合。在$t$年 5 月，依靠 A 股所有股票的 5 个规模分位点以及 5 个账面市值比分位点，构建起 25 个投资组合，并计算各投资组合从$t$年 5 月至$t+1$年 4 月的月度加权平均收益率。**这 25 个投资组合在 2000 年 1 月至 2019 年 12 月中的月度加权平均收益率是时间序列回归中的被解释变量。**

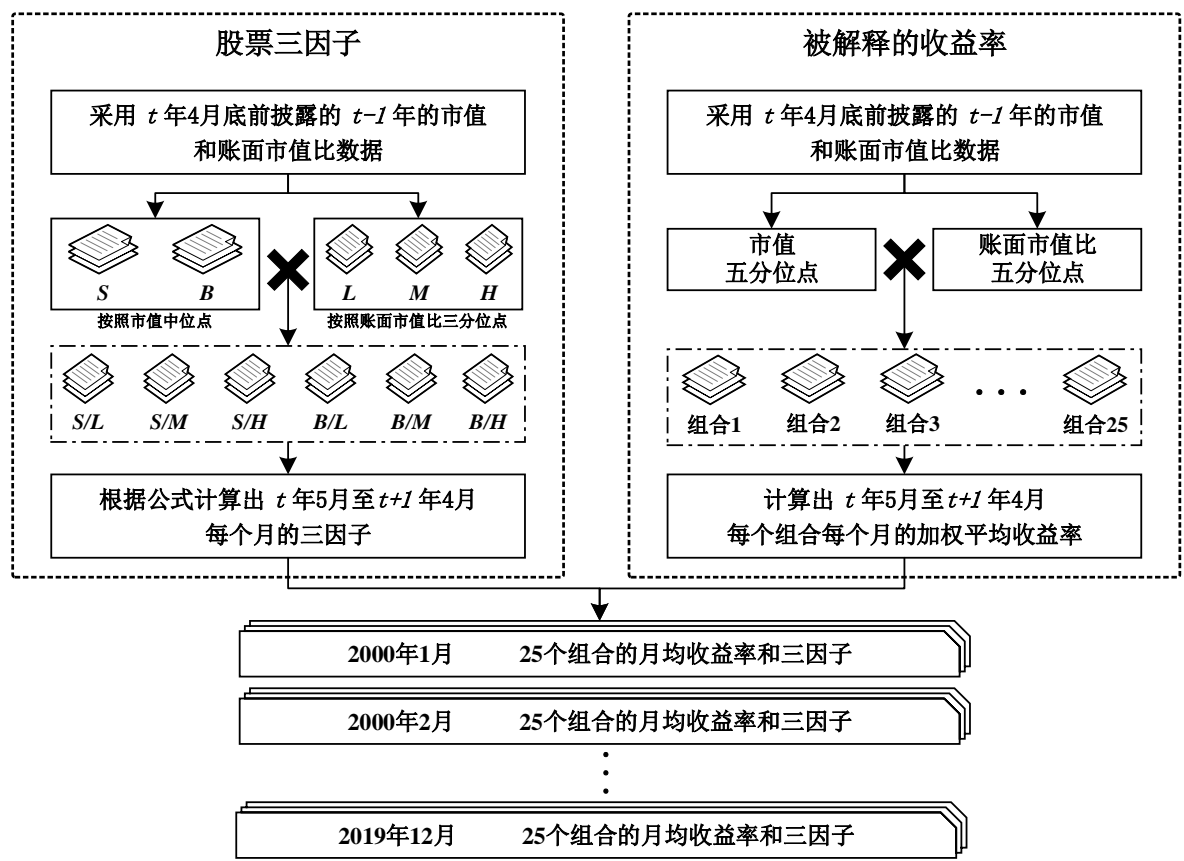

**图 3.** Fama-French 三因子模型的时间序列回归数据构建框架图

### 4.2.3 数据处理结果展示

图 3.对上述股票三因子以及收益率的构造过程做了简洁总结。表 5.则呈现了处理后的时间序列数据结构。

表 5. 时间序列数据结构

| 组合 | 时间 | 被解释收益率 | 股票三因子 | | |
|---|---|---|---|---|---|
| | | | $RM-RF$ | $SMB$ | $HML$ |
| 组合 1 | 2000 年 1 月 | $R_1(1)-RF(1)$ | $RM(1)-RF(1)$ | $SMB(1)$ | $HML(1)$ |
| | ⋮ | ⋮ | ⋮ | ⋮ | ⋮ |
| | 2000 年 12 月 | $R_1(12)-RF(12)$ | $RM(12)-RF(12)$ | $SMB(12)$ | $HML(12)$ |
| | ⋮ | ⋮ | ⋮ | ⋮ | ⋮ |
| | 2019 年 1 月 | $R_1(229)-RF(229)$ | $RM(229)-RF(229)$ | $SMB(229)$ | $HML(229)$ |
| | ⋮ | ⋮ | ⋮ | ⋮ | ⋮ |
| | 2019 年 12 月 | $R_1(240)-RF(240)$ | $RM(240)-RF(240)$ | $SMB(240)$ | $HML(240)$ |
| ⋮ | ⋮ | ⋮ | ⋮ | ⋮ | ⋮ |
| 组合 20 | 2000 年 1 月 | $R_{25}(1)-RF(1)$ | $RM(1)-RF(1)$ | $SMB(1)$ | $HML(1)$ |
| | ⋮ | ⋮ | ⋮ | ⋮ | ⋮ |
| | 2000 年 12 月 | $R_{25}(12)-RF(12)$ | $RM(12)-RF(12)$ | $SMB(12)$ | $HML(12)$ |
| | ⋮ | ⋮ | ⋮ | ⋮ | ⋮ |
| | 2019 年 1 月 | $R_{25}(229)-RF(229)$ | $RM(229)-RF(229)$ | $SMB(229)$ | $HML(229)$ |
| | ⋮ | ⋮ | ⋮ | ⋮ | ⋮ |
| | 2019 年 12 月 | $R_{25}(240)-RF(240)$ | $RM(240)-RF(240)$ | $SMB(240)$ | $HML(240)$ |

### 4.2.4 数据的初步描述性统计

表 6.展示了本文所用交易数据根据规模和账面市值比分位点构建的 25 个投资组合基本统计信息。可以看出，表中的结果和 Fama 和 French（1993）基于纽约证券交易所分位点形成投资组合的描述性统计有所不同，公司数量和规模大小以及账面市值比之间**没有明显的关系**。正如所预想的那样，5 个最大规模的投资组合的市值约为总市值的 45.97%；最大市值、最低账面市值比的股票占据了总市值的 16%以上。

表 **6.** 基于规模和账面市值比构建的 25 个投资组合的描述性统计（2000 至 2019 年，20 年）

| 规模 | 账面市值比 | | | | | | | | | |
|---|---|---|---|---|---|---|---|---|---|---|
| | Low | 2 | 3 | 4 | High | Low | 2 | 3 | 4 | High |
| | 平均每个月的市值 | | | | | 平均每个月的账面市值比 | | | | |
| Small | 9.3 | 9.0 | 8.3 | 7.7 | 7.2 | .34 | .58 | .79 | 1.08 | 2.24 |
| 2 | 14.8 | 14.8 | 14.7 | 14.8 | 14.6 | .35 | .57 | .79 | 1.06 | 1.95 |
| 3 | 22.7 | 22.8 | 22.6 | 22.5 | 22.5 | .35 | .58 | .68 | 1.06 | 1.99 |
| 4 | 37.6 | 37.2 | 37.4 | 36.9 | 37.3 | .35 | .57 | .78 | 1.06 | 2.11 |
| Big | 138.4 | 115.6 | 157.2 | 190.1 | 204.5 | .32 | .57 | .78 | 1.07 | 2.35 |
| | 平均每个月的总市值占比 | | | | | 平均每个月的公司数 | | | | |
| Small | .0059 | .0082 | .0087 | .0100 | .0100 | 30.1 | 46.3 | 60.3 | 90.3 | 111.1 |
| 2 | .0123 | .0153 | .0167 | .0155 | .0125 | 60.9 | 70.3 | 75.1 | 72.4 | 60.9 |
| 3 | .0197 | .0228 | .0232 | .0226 | .0184 | 67.7 | 74.9 | 76.1 | 69.6 | 53.0 |
| 4 | .0373 | .0382 | .0350 | .0311 | .0275 | 77.7 | 82.9 | 72.1 | 61.0 | 49.7 |
| Big | .1688 | .0921 | .1017 | .0971 | .1496 | 98.1 | 67.7 | 59.2 | 50.9 | 69.2 |

如表 7.所示，基于规模和账面市值比构建的 25 个股票组合的月平均超额收益率范围较广，从每月 0.49%到 2.06%。基于本表，可以得出与 Fama 和 French（1992a）部分相同的结论：一方面，按照账面市值比划分的五组中，**平均收益率均随着规模增大而减少**，且规模最大组和最小组之间的平均收益率差异也是较为显著的；但另一方面，按照规模划分的五组中，平均回报率**没有呈现**"随着账面市值比增加而增加的趋势"，最高价值组和最低价值组收益率之差并不显著为正，甚至在 Small 和 2 组中为负数，这是匪夷所思的，本文将会在后面进行补充讨论。

表 **7.** 基于规模和账面市值比构建的 25 个投资组合的月平均超额收益率

| 规模 | 账面市值比 | | | | | |
|---|---|---|---|---|---|---|
| | Low | 2 | 3 | 4 | High | High-Low |
| Small | 1.96 | 2.06 | 1.91 | 1.84 | 1.50 | -0.459 |
| 2 | 1.44 | 1.59 | 1.57 | 1.61 | 1.28 | -0.158 |
| 3 | 0.85 | 1.07 | 1.38 | 1.22 | 1.14 | 0.286 |
| 4 | 0.76 | 0.84 | 1.00 | 1.03 | 1.01 | 0.252 |
| Big | 0.49 | 0.76 | 0.69 | 0.78 | 0.79 | 0.295 |
| Big-Small | -1.467*** | -1.305*** | -1.226*** | -1.056** | -0.713 | |

$^{*}\ p < 0.1$, $^{**}\ p < 0.05$, $^{***}\ p < 0.01$

基于中国 A 股交易数据的资本资产定价模型实证研究　　　　　　　　　　　　　　　　　任凯## 4.3 时间序列回归结果

在时间序列回归中，斜率和 $R^2$ 值是不同因子是否捕捉到股票收益变化的直接证据。本文和 Fama 和 French（1993）一样，通过三种回归来验证三因子在股票收益中发挥的作用，分别是：

（a）只使用市场因子即超额市场收益 $RM - RF$ 进行回归；

（b）使用市值因子 $SMB$ 和价值因子 $HML$ 作为解释变量的回归；

（c）使用三因子 $RM - RF$、$SMB$ 和 $HML$ 进行回归。

本节将报告上述三种回归的结果，分析三因子对 A 股市场收益率的解释能力。

### 4.3.1 市场因子

表 8.显示超额市场收益 $RM - RF$ 一定程度上能解释股票收益率的变化，但**不够充分**。25 个组合中只有 4 个较高市值组合的调整 $R^2$ 超过了 0.9，小市值股票组合的调整 $R^2$ 大都在 0.7 至 0.9 之间，规模最小的部分组合甚至低于了 0.7。

**表 8.** 25 个组合的超额股票收益率对市场因子 $RM - RF$ 的回归结果

（2000 年 1 月至 2019 年 12 月，240 个月）

$$R(t) - RF(t) = a + b[RM(t) - RF(t)] + e(t)$$

| | 相关回归参数值 | | | | | | | | | |
|---|---|---|---|---|---|---|---|---|---|---|
| | 账面市值比 | | | | | | | | | |
| 规模 | Low | 2 | 3 | 4 | High | Low | 2 | 3 | 4 | High |
| | *b* | | | | | *t(b)* | | | | |
| Small | 1.13*** | 1.11*** | 1.13*** | 1.14*** | 1.10*** | 20.39 | 18.91 | 21.97 | 19.37 | 20.98 |
| 2 | 1.10*** | 1.12*** | 1.11*** | 1.13*** | 1.12*** | 19.23 | 21.36 | 22.60 | 24.46 | 20.32 |
| 3 | 1.11*** | 1.09*** | 1.11*** | 1.13*** | 1.12*** | 25.37 | 29.44 | 26.03 | 25.68 | 30.26 |
| 4 | 1.06*** | 1.10*** | 1.12*** | 1.15*** | 1.09*** | 24.16 | 26.46 | 33.22 | 33.10 | 45.13 |
| Big | 0.95*** | 1.04*** | 1.04*** | 1.00*** | 0.94*** | 33.45 | 50.02 | 40.32 | 40.82 | 25.40 |
| | 调整的 $R^2$ | | | | | *s(e)* | | | | |
| Small | 0.65 | 0.68 | 0.69 | 0.70 | 0.73 | 6.66 | 6.21 | 6.10 | 6.04 | 5.45 |
| 2 | 0.73 | 0.76 | 0.76 | 0.79 | 0.81 | 5.34 | 5.13 | 5.01 | 4.63 | 4.42 |
| 3 | 0.78 | 0.81 | 0.82 | 0.84 | 0.88 | 4.77 | 4.26 | 4.18 | 3.99 | 3.30 |
| 4 | 0.79 | 0.85 | 0.89 | 0.90 | 0.92 | 4.35 | 3.75 | 3.25 | 3.14 | 2.57 |
| Big | 0.85 | 0.93 | 0.94 | 0.90 | 0.83 | 3.28 | 2.24 | 2.20 | 2.69 | 3.43 |

*  $p < 0.1$, **  $p < 0.05$, ***  $p < 0.01$

### 4.3.2 规模因子和价值因子

表 9.显示,在没有市场因子的情况下,$SMB$ 和 $HML$ 也能**捕捉到股票收益的少量变化**:25 个调整 $R^2$ 值中有 20 个高于 0.1,12 个高于 0.3。然而对于规模最大的五分之一的投资组合,$SMB$ 和 $HML$ 对股票收益的解释能力并不强,这似乎刚好和表 8.形成了"互补",表 8.中调整 $R^2$ 较低的几个组合在表 9.中的调整 $R^2$ 也明显较高。

表 9. 25 个组合的超额股票收益率对规模因子($SMB$)和价值因子($HML$)的回归结果
(2000 年 1 月至 2019 年 12 月,240 个月)

$$R(t) - RF(t) = sSMB(t) + hHML(t) + e(t)$$

| | 相关回归参数值 | | | | | | | | | |
|---|---|---|---|---|---|---|---|---|---|---|
| | 账面市值比 | | | | | | | | | |
| 规模 | Low | 2 | 3 | 4 | High | Low | 2 | 3 | 4 | High |
| | | | $s$ | | | | | $t(s)$ | | |
| Small | 1.63*** | 1.61*** | 1.76*** | 1.80*** | 1.71*** | 7.62 | 8.53 | 10.24 | 9.34 | 9.10 |
| 2 | 1.48*** | 1.50*** | 1.52*** | 1.51*** | 1.52*** | 8.17 | 9.18 | 8.50 | 8.78 | 8.03 |
| 3 | 1.25*** | 1.27*** | 1.30*** | 1.35*** | 1.19*** | 7.31 | 7.57 | 7.99 | 7.77 | 7.00 |
| 4 | 1.05*** | 1.13*** | 1.05*** | 1.07*** | 0.92*** | 6.87 | 6.40 | 5.50 | 5.40 | 5.20 |
| Big | 0.24 | 0.38** | 0.48** | 0.42** | 0.13 | 1.47 | 2.27 | 2.60 | 2.38 | 0.89 |
| | | | $h$ | | | | | $t(h)$ | | |
| Small | 0.08 | -0.07 | 0.16 | 0.32* | 0.50 | 0.29 | -0.21 | 0.61 | 1.05 | 1.90 |
| 2 | 0.09 | 0.09 | 0.19 | 0.44** | 0.76 | 0.33 | 0.34 | 0.67 | 1.59 | 2.49 |
| 3 | -0.15 | 0.10 | 0.23* | 0.54*** | 0.76 | -0.51 | 0.37 | 0.93 | 1.84 | 2.71 |
| 4 | -0.26 | 0.11 | 0.30* | 0.45*** | 0.83 | -1.00 | 0.38 | 1.13 | 1.68 | 3.04 |
| Big | -0.68** | -0.08 | 0.49*** | 0.81*** | 0.92 | -2.48 | -0.30 | 1.59 | 2.74 | 3.67 |
| | | | 调整的 $R^2$ | | | | | $s(e)$ | | |
| Small | 0.37 | 0.41 | 0.45 | 0.44 | 0.41 | 8.99 | 8.37 | 8.16 | 8.24 | 8.00 |
| 2 | 0.36 | 0.37 | 0.37 | 0.34 | 0.32 | 8.25 | 8.23 | 8.16 | 8.31 | 8.31 |
| 3 | 0.31 | 0.29 | 0.29 | 0.27 | 0.21 | 8.43 | 8.23 | 8.34 | 8.51 | 8.58 |
| 4 | 0.27 | 0.23 | 0.18 | 0.17 | 0.13 | 8.20 | 8.45 | 8.68 | 8.91 | 8.57 |
| Big | 0.10 | 0.03 | 0.03 | 0.05 | 0.08 | 7.94 | 8.55 | 8.55 | 8.29 | 7.97 |

*$p < 0.1$, **$p < 0.05$, ***$p < 0.01$

### 4.3.3 三因子

表 9.显示，单独使用 $SMB$ 和 $HML$ 对股票收益的**解释力并不强**，但似乎与市场因子形成**互补**。表 10.展示了将三个因子全部包含在模型后的回归结果，不难发现：**在本文的数据上，Fama-French 三因子模型得到了较好的验证，这三个因子的确能够解释股票收益的大部分变化。**

首先，$SMB$ 因子斜率的 $t$ 统计量几乎均大于 2，大多数大于 10，这说明 $SMB$ 作为对规模风险的测度，显著地捕捉到了市场因子和 $HML$ 所忽略的股票收益率变化。此外，$SMB$ 的斜率与规模大小明显相关，在每一个账面市值比的五分位数分组中，$SMB$ 的斜率从较小规模到较大规模均呈现单调递减的趋势。

其次，$HML$ 的斜率，即对价值风险回报的测度与账面市值比之间也有相关关系。在每个规模的五分之一分组中，$HML$ 的斜率单调地增加，从最低账面市值比五分之一的负值到最高五分之一的正值，且几乎所有的值都在 1%的显著性水平下都是显著的。$HML$ 明显捕捉到了市场因子和 $SMB$ 所忽略的股票收益的变化。

最后，三因子模型中调整的 $R^2$ 大幅上升。表 8.中单独使用市场因子回归，只有四个组合的调整 $R^2$ 大于 0.9，而在表 10.中小于 0.9 的调整 $R^2$ 则成了"凤毛麟角"（25 个中只有 1 个小于 0.9）。对于规模最小的五个投资组合，调整的 $R^2$ 从表 8.中 0.65 到 0.73 显著提升为了表 10.中的 0.87 到 0.96。在表 10.中即使是最小的 $R^2$（最小规模和最小账面市值比组合的 0.88），也远远大于市场因子单独产生的 0.65。

另外，在表 8.的单因子回归中，最小规模和最低账面市值比组合的 $\beta$ 是 1.13，最大规模和最高账面市值比组合的 $\beta$ 是 0.94，而在表 10.的三因子回归中，这两个组合的 $\beta$ 分别是 1.02 和 0.97，这说明在回归中加入 $SMB$ 和 $HML$ 会使股票的 $\beta$ 向 1.0 塌陷：低 $\beta$ 向 1.0 移动，高 $\beta$ 则向下移动，一定程度上修正了 $\beta$ 因子对风险的度量偏差。

## 4.4 对 Fama-French 三因子模型验证的结论

本文在表 8.、表 9.、表 10.中的结果充分说明对 Fama-French 三因子模型对 2000 年 1 月至 2019 年 12 月 A 股市场中的交易数据有较强的解释性，规模因子、价值因子能够显著地捕捉到收益率的变动。

**表 10.** 25 个组合的超额股票收益率对三因子的回归结果
（2000 年 1 月至 2019 年 12 月，240 个月）
$$R(t)-RF(t)=a+b[RM(t)-RF(t)]+sSMB(t)+hHML(t)+e(t)$$

| | 相关回归参数值 | | | | | | | | | |
|---|---|---|---|---|---|---|---|---|---|---|
| | 账面市值比 | | | | | | | | | |
| 规模 | Low | 2 | 3 | 4 | High | Low | 2 | 3 | 4 | High |
| | $b$ | | | | | $t(b)$ | | | | |
| Small | 1.02*** | 1.00*** | 1.00*** | 1.01*** | 0.98*** | 27.71 | 38.64 | 57.32 | 60.37 | 44.19 |
| 2 | 0.99*** | 1.01*** | 1.01*** | 1.02*** | 1.01*** | 56.27 | 59.67 | 59.57 | 52.64 | 41.08 |
| 3 | 1.03*** | 1.01*** | 1.02*** | 1.04*** | 1.05*** | 40.88 | 55.52 | 60.05 | 44.14 | 52.33 |
| 4 | 1.00*** | 1.03*** | 1.06*** | 1.09*** | 1.05*** | 44.09 | 53.40 | 49.09 | 54.58 | 60.45 |
| Big | 0.97*** | 1.05*** | 1.04*** | 1.01*** | 0.97*** | 49.99 | 59.34 | 41.64 | 57.55 | 38.01 |
| | $s$ | | | | | $t(s)$ | | | | |
| Small | 1.14*** | 1.13*** | 1.28*** | 1.32*** | 1.24*** | 14.48 | 18.30 | 36.14 | 25.45 | 21.92 |
| 2 | 1.00*** | 1.02*** | 1.04*** | 1.02*** | 1.03*** | 15.65 | 23.27 | 32.05 | 20.95 | 21.13 |
| 3 | 0.76*** | 0.79*** | 0.81*** | 0.85*** | 0.69*** | 12.75 | 18.38 | 18.96 | 19.35 | 13.94 |
| 4 | 0.57*** | 0.63*** | 0.54*** | 0.55*** | 0.42*** | 9.76 | 13.3 | 9.88 | 8.87 | 10.07 |
| Big | -0.22*** | -0.12*** | -0.03 | -0.07 | -0.33*** | -4.59 | -3.21 | -0.49 | -1.32 | -7.74 |
| | $h$ | | | | | $t(h)$ | | | | |
| Small | -0.27*** | -0.41*** | -0.18*** | -0.02 | 0.16* | -3.42 | -6.33 | -3.43 | -0.42 | 1.80 |
| 2 | -0.25** | -0.26*** | -0.15*** | 0.09 | 0.41*** | -2.32 | -4.42 | -3.33 | 1.15 | 5.37 |
| 3 | -0.5*** | -0.25*** | -0.13* | 0.19** | 0.40*** | -6.02 | -4.21 | -1.77 | 2.31 | 5.75 |
| 4 | -0.6*** | -0.25*** | -0.07 | 0.08 | 0.47*** | -7.09 | -3.55 | -0.96 | 0.99 | 9.28 |
| Big | -1.02*** | -0.45*** | 0.13 | 0.46*** | 0.59*** | -13.96 | -7.29 | 1.47 | 5.24 | 9.07 |
| | 调整的 $R^2$ | | | | | $s(e)$ | | | | |
| Small | 0.88 | 0.94 | 0.97 | 0.96 | 0.96 | 4.00 | 2.76 | 1.98 | 2.09 | 2.05 |
| 2 | 0.94 | 0.97 | 0.97 | 0.97 | 0.96 | 2.50 | 1.79 | 1.76 | 1.90 | 2.08 |
| 3 | 0.95 | 0.96 | 0.96 | 0.95 | 0.95 | 2.24 | 1.91 | 1.99 | 2.13 | 2.15 |
| 4 | 0.94 | 0.95 | 0.95 | 0.95 | 0.95 | 2.25 | 2.05 | 2.18 | 2.22 | 2.03 |
| Big | 0.95 | 0.95 | 0.94 | 0.93 | 0.94 | 1.96 | 1.94 | 2.16 | 2.21 | 2.02 |

\* $p<0.1$, \*\* $p<0.05$, \*\*\* $p<0.01$

# 五、 中国版三因子模型

和大多数基于中国股市数据研究三因子模型的文章一样，本文的结果也存在一定的异常。上文提及的表 7.中体现的"平均回报率没有呈现随着账面市值比的增加而增加的趋势"这一结果让笔者进行了更加深入的思考。

刘等人（2019）认为规模因子旨在捕捉与规模相关的风险差异，这些差异来自于相关业务，然而中国小型上市企业的股票价格通常反映了大量与企业基础业务无关的信息，而是与首次公开发行（IPO）相关。在中国，部分私营企业因于 IPO 市场的严格监管选择通过反向兼并及时上市，即私营企业将一家上市公司作为壳目标，通过收购其股份获得控制权，壳公司购买该企业的资产以换取新发行的股票，这导致了**"壳价值污染现象"**。

中国规模最小的 30%的股票大多涉及到反向并购，这些公司的收益与经营基本面的关系较小而与 IPO 活动的关系较大。为了避免在构建因子时受到"壳价值污染现象"的不利影响，本文选择剔除市值最低 30%部分的股票后得到了表 11.中的结果。与表 7.相比，该表中的结果明显能够更好地支持"平均回报率随着账面市值比的增加而增加的"这一结论，最高价值组和最低价值组平均收益率之差没有出现负值，而且部分正值在 10%的显著性水平上是显著的。

表 11. 基于规模和账面市值比构建的 25 个投资组合的月平均超额收益率
（去除市值底部 30%的股票后）

| 规模 | 账面市值比 | | | | | |
|---|---|---|---|---|---|---|
| | Low | 2 | 3 | 4 | High | High-Low |
| Small | 1.17 | 1.43 | 1.49 | 1.53 | 1.20 | 0.035 |
| 2 | 0.74 | 1.13 | 1.32 | 1.30 | 1.19 | 0.448* |
| 3 | 0.69 | 0.98 | 1.10 | 1.15 | 0.64 | 0.464* |
| 4 | 0.64 | 0.90 | 1.00 | 1.08 | 0.99 | 0.352 |
| Big | 0.53 | 0.61 | 0.77 | 0.75 | 0.75 | 0.220 |
| Big-Small | -0.638 | -0.815** | -0.724** | -0.789** | -0.454 | |

* $p < 0.1$, ** $p < 0.05$, *** $p < 0.01$

另外本文也进一步回归得到了表12.中的结果，和表10.相比，可以发现调整 $R^2$ 由最低的0.88提升至了0.93。这都说明了在中国A股市场中，采用剔除小规模股票的方式来减少"壳价值污染"的不利影响是可行的。反过来，若想在中国市场中更为有效地使用三因子模型必须要考虑到该影响，并尽量剔除。

**表 12.** 25 个组合的超额股票收益率对三因子的回归结果（去除市值底部 30%的股票后）
（2000 年 1 月至 2019 年 12 月，240 个月）
$$R(t) - RF(t) = a + b[RM(t) - RF(t)] + sSMB(t) + hHML(t) + e(t)$$

| | 相关回归参数值 | | | | | | | | | |
|---|---|---|---|---|---|---|---|---|---|---|
| | 账面市值比 | | | | | | | | | |
| Size | Low | 2 | 3 | 4 | High | Low | 2 | 3 | 4 | High |
| | b | | | | | t(b) | | | | |
| Small | 0.99*** | 0.98*** | 1.00*** | 1.00*** | 1.01*** | 36.21 | 39.52 | 55.92 | 48.97 | 41.98 |
| 2 | 1.01*** | 1.00*** | 0.99*** | 1.02*** | 1.01*** | 32.93 | 55.95 | 51.61 | 50.58 | 55.98 |
| 3 | 0.98*** | 0.99*** | 1.03*** | 1.05*** | 1.03*** | 42.33 | 55.85 | 57.34 | 55.05 | 52.68 |
| 4 | 0.96*** | 1.06*** | 1.05*** | 1.08*** | 1.06*** | 34.56 | 48.79 | 45.44 | 51.70 | 49.66 |
| Big | 0.99*** | 1.03*** | 1.07*** | 1.01*** | 0.95*** | 42.12 | 53.68 | 36.81 | 39.02 | 40.35 |
| | s | | | | | t(s) | | | | |
| Small | 1.06*** | 1.07*** | 1.12*** | 1.06*** | 1.06*** | 19.42 | 19.16 | 27.14 | 21.07 | 18.16 |
| 2 | 0.97*** | 0.89*** | 0.87*** | 1.05*** | 0.84*** | 16.37 | 15.54 | 12.82 | 21.98 | 21.79 |
| 3 | 0.71*** | 0.83*** | 0.74*** | 0.81*** | 0.64*** | 12.10 | 15.80 | 17.11 | 15.18 | 12.25 |
| 4 | 0.45*** | 0.51*** | 0.53*** | 0.55*** | 0.38*** | 7.15 | 7.77 | 8.59 | 10.52 | 7.35 |
| Big | -0.35*** | -0.22*** | -0.05 | -0.12 | -0.41*** | -5.75 | -4.30 | -0.51 | -1.50 | -8.25 |
| | h | | | | | t(h) | | | | |
| Small | -0.33*** | -0.27*** | -0.16*** | -0.07 | 0.25*** | -3.80 | -5.18 | -4.13 | -1.41 | 4.00 |
| 2 | -0.42*** | -0.30*** | -0.17** | 0.08 | 0.34*** | -6.85 | -4.88 | -2.28 | 1.50 | 7.31 |
| 3 | -0.63*** | -0.35*** | -0.13** | 0.02 | 0.35*** | -9.31 | -6.19 | -2.53 | 0.41 | 7.63 |
| 4 | -0.73*** | -0.33*** | -0.1 | 0.19*** | 0.45*** | -11.31 | -4.62 | -1.64 | 3.74 | 8.29 |
| Big | -0.97*** | -0.42*** | 0.06 | 0.34*** | 0.54*** | -14.56 | -7.82 | 0.58 | 4.33 | 8.60 |
| | 调整的 $R^2$ | | | | | s(e) | | | | |
| Small | 0.93 | 0.96 | 0.97 | 0.97 | 0.96 | 2.76 | 2.01 | 1.82 | 1.89 | 2.13 |
| 2 | 0.94 | 0.96 | 0.95 | 0.96 | 0.96 | 2.43 | 2.05 | 2.08 | 1.92 | 1.87 |
| 3 | 0.95 | 0.96 | 0.96 | 0.96 | 0.95 | 2.19 | 1.96 | 2.01 | 1.98 | 2.09 |
| 4 | 0.93 | 0.94 | 0.94 | 0.95 | 0.95 | 2.41 | 2.41 | 2.26 | 2.17 | 2.04 |
| Big | 0.94 | 0.93 | 0.92 | 0.91 | 0.94 | 2.07 | 2.24 | 2.54 | 2.58 | 2.00 |

* $p < 0.1$, ** $p < 0.05$, *** $p < 0.01$

# 六、 结论

本文基于中国 A 股市场 2000 年到 2019 年的交易数据，对资本资产定价模型进行实证研究。整体上分为对标准 CAPM 的检验和对 Fama-French 三因子模型的分析。

首先，对标准 CAPM 的检验结果可以使本文得到如下结论：1）投资组合风险和其预期收益之间的关系的确是**线性**的；2）**不能拒绝**除了投资组合风险外，没有任何风险措施会系统地影响平均收益这一假设；3）无风险利率在标准 CAPM 模型**近似充当常数项**的作用。但是"收益和风险之间有正向的关系"这一重要结论并未被证实。其次，对 Fama-French 三因子模型的分析结果充分说明了：对于我国该阶段的股票交易数据而言，市场因子、规模因子 *SMB* 和价值因子 *HML* **能够有效地解释股票收益率变化**，但依然存在部分异常。最后，本文通过剔除小规模 30%股票的方式尽量减弱了壳价值污染的不利影响，对异常现象进行了一定程度的解释。

本文最大的亮点在于：选取了中国市场交易数据进行研究，详细阐释了资本资产定价模型在中国市场中的适用情况，且对两模型的具体实证研究方法做了细致的呈现。

# 参 考 文 献